\documentclass[reprint,amsmath,amssymb,prl,longbibliography]{revtex4-1}

\usepackage{graphicx}
\usepackage[utf8]{inputenc}
\usepackage{color}
\definecolor{goodgreen}{rgb}{0.1,0.5,0}
\definecolor{goodred}{rgb}{0.7,0,0}
\usepackage[colorlinks,urlcolor=goodgreen,citecolor=blue,linkcolor=goodred]%
{hyperref}
\usepackage{upgreek}

\newcommand{\vg}{\ensuremath{V_\text{g}}}
\newcommand{\cg}{\ensuremath{C_\text{g}}}

\newcommand{\ctot}{\ensuremath{C_{\Sigma}}}
\newcommand{\vsd}{\ensuremath{V_\text{sd}}}

\newcommand{\un}[1]{\ensuremath{\,\text{#1}}}
\newcommand{\nel}{\ensuremath{N}}

\begin{document}

\title{Nanomechanical characterization of the Kondo charge dynamics in 
a carbon nanotube}

\author{K. J. G. Götz}
\author{D. R. Schmid}
\author{F. J. Schupp}
\author{P. L. Stiller}
\author{Ch. Strunk}
\author{A. K. Hüttel}
\email{andreas.huettel@ur.de}
\affiliation{Institute for Experimental and Applied Physics, 
University of Regensburg, Universitätsstr.\ 31, 93053 Regensburg, 
Germany
}

\date{\today}

\begin{abstract}
Using the transversal vibration resonance of a suspended carbon nanotube as
charge detector for its embedded quantum dot, we investigate the case of strong 
Kondo correlations between a quantum dot and its leads. We demonstrate that 
even when large Kondo conductance is carried at odd electron number, the 
charging behaviour remains similar between odd and even quantum dot occupation. 
While the Kondo conductance is caused by higher order processes, a sequential 
tunneling only model can describe the time-averaged charge. The gate potentials 
of maximum current and fastest charge increase display a characteristic 
relative shift, which is suppressed at increased temperature. These
observations agree very well with models for Kondo-correlated quantum dots.
\end{abstract}

\maketitle 

The Kondo effect \cite{Kondo1964} is a striking manifestation of electronic
correlations. In semiconductor quantum dots as Coulomb blockade systems
\cite{beenakker}, in its most prevalent type it expresses itself as a distinct
zero-bias maximum of differential conductance at odd electronic occupation
\cite{prl-goldhaber:5225, nature-goldhaber:156, Kouwenhoven1998}. In spite of
this strong impact on electronic {\em charge} transport, the degeneracy central
to its formation is then given by the {\em spin} states of an unpaired
electron: for the SU(2) spin Kondo effect, below a characteristic temperature
$T_K$, exchange coupling between a localized electron and conduction band
charges leads to the formation of the Kondo resonance at the Fermi level.
A question that lends itself immediately is how the strongly enhanced Kondo
conductance within Coulomb blockade relates to the precise charge trapped
within the quantum dot and its evolution as a function of applied gate voltage
\cite{prl-gerland:3710, prl-sprinzak:176805, Desjardins2017}.

Suspended carbon nanotube quantum dots provide extraordinarily clean and
controllable mesoscopic model systems \cite{nmat-cao:745, rmp-laird}, where
transport spectra from single- and few electron physics
\cite{nphys-deshpande:314, nature-kuemmeth:448, pecker:natphys2013,
magda} all the way to open systems and electronic Fabry-Perot interferometry
\cite{nature-liang:665, prl-dirnaichner} can be analyzed. Also regarding Kondo
phenomena a wide range of experimental work on carbon nanotubes exists
\cite{nature-nygaard:342, nature-jarillo:484, prl-makarovski:066801, kondoso,
niklas:natcomms2016}, making use of the well-characterized electronic structure.
Then again, as nano-electromechanical systems, carbon nanotubes have shown at
cryogenic temperatures exceedingly high mechanical quality factors \cite{highq,
pssb-huettel, mosermillion} and strong interaction between single electron
tunneling and vibrational motion \cite{strongcoupling, science-lassagne:1107,
magdamping, prb-meerwaldt:115454}. The detection of the transversal vibration
frequency of a carbon nanotube provides a powerful means to measure the charge
on its embedded quantum dot \cite{strongcoupling, prb-meerwaldt:115454}. 

In this Letter, we investigate the parameter region of strong Kondo
correlations between a suspended nanotube quantum dot and its metallic leads
\cite{prl-goldhaber:5225, nature-goldhaber:156, Kouwenhoven1998, 
nature-nygaard:342, nature-jarillo:484, prl-makarovski:066801, kondoso,
niklas:natcomms2016, kiselevprl2013}. We measure the gate voltage dependence of 
the time-averaged charge $e \left< N \right> \! (\vg)$ on the quantum dot. The 
observed typical asymmetry in {\em conductance} between odd and even occupation 
states, indicating SU(2) Kondo behavior, is clearly absent in the 
gate-dependent trapped {\em charge}. This shows that the current is carried by 
higher-order processes leading to {\em only virtual} occupation on the quantum 
dot, while the time-averaged charge remains determined by the first-order 
processes of sequential tunneling. In addition, we observe a distinct gate 
voltage offset between charging of the quantum dot and the current maximum, 
which is suppressed at increasing temperature. Our results agree very well with
theoretical studies of Kondo-correlated quantum dots \cite{prl-gerland:3710,
Wingreen1994}.

\begin{figure}[t]
\includegraphics[width=\columnwidth]{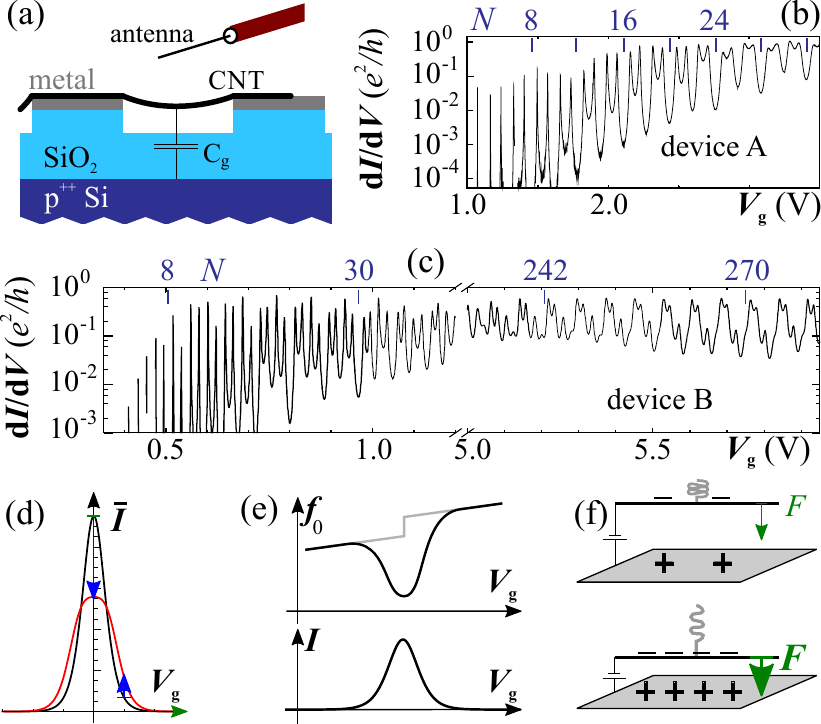}
\caption{\label{fig-gatetrace} 
(a) Sketch of the device geometry (not to scale). See the Supplement
\cite{supplement} for a table of the device properties.
(b,c) Low-bias differential conductance $\text{d}I / \text{d}\vsd$ of carbon 
nanotube devices A and B as a function of applied back gate voltage \vg. $N$
indicates the number of trapped electrons. The transition from strong Coulomb
blockade (left edge) to strongly Kondo enhanced transport is visible in both
cases.
(d) At resonant driving a nanotube vibrates strongly, leading to a fast
oscillation of $C_g$ and, averaged over the vibration, a broadening of
Coulomb oscillations.
(e) Typical gate voltage evolution of the transversal vibration resonance
frequency and the current in the Coulomb blockade regime, see
\cite{strongcoupling} and the text.
(f) Principle of electrostatic vibration softening: when a vibrating capacitor 
at constant voltage adapts its charge to the momentary position, an 
electrostatic force opposite to the mechanical restoring force occurs. This 
results in an effective smaller spring constant and resonance frequency.
}
\end{figure}
\paragraph{Device characterization--} 
Figure~\ref{fig-gatetrace}(a) displays a sketch of our device structure; a
table of fabrication parameters can be found in the Supplement 
\cite{supplement}. On a highly $\text{p}^{++}$ doped Si substrate with
thermally grown $\text{SiO}_2$ on top, electrode patterns were defined via
electron beam lithography and metal evaporation. The metal layer directly
serves as etch mask for subsequent anisotropic dry etching of the oxide,
generating deep trenches between the electrodes. As last steps, growth catalyst
was locally deposited and carbon nanotubes were grown via chemical vapor
deposition \cite{nature-kong:878}.

Electronic transport measurements were performed in a dilution refrigerator at 
$T_\text{MC} \le 25\un{mK}$. The measurement setup combines dc current
measurement as required for Coulomb blockade transport spectroscopy
\cite{kouwenhoven} with radio-frequency irradiation using an antenna several 
millimeters from the device \cite{highq, strongcoupling, pssb-huettel}. As can 
be seen from the differential conductance in linear response in
Fig.~\ref{fig-gatetrace}(b,c), in both devices close to a small band gap
Coulomb blockade and sequential tunneling dominates. For larger positive gate
voltages \vg\ the transparency of the tunneling barriers increases. This leads 
to a crossover towards regular Kondo enhancement of the conductance 
\cite{nature-goldhaber:156, nature-nygaard:342}; the clear two-fold pattern in 
Fig.~\ref{fig-gatetrace}(b) indicates approximate SU(2) Kondo behaviour 
\footnote{The maximum conductance value lower than $2e^2/h$ is likely due to 
asymmetric tunnel barrier transparencies \cite{prl-ng-1988}. A larger version of
Fig.~\ref{fig-gatetrace}(b,c) can be found in the Supplement \cite{supplement}
as Fig.~S1.}.

\paragraph{Mechanical resonance detection--}
With a radio-frequency signal applied at the mechanical resonance $f_0$ of its 
transversal vibration mode, the nanotube is driven into motion, leading to a
change in detected dc current \cite{highq, strongcoupling, pssb-huettel,
prb-meerwaldt:115454}: the capacitance between back gate and nanotube $\cg$ is
modulated with the deflection, broadening the Coulomb oscillations in a slow dc
measurement, as shown in Fig.~\ref{fig-gatetrace}(d). This enables us to detect
$f_0$ and its dependence on the back gate voltage \vg\ in the dc-current.

Figure~\ref{fig-gatetrace}(e) sketches a typical evolution of the resonance 
frequency with increasing positive gate voltage in the strong Coulomb blockade
regime \cite{strongcoupling}. The continuous increase of the gate charge and
the discrete increase of the quantum dot charge both contribute via mechanical
tension to $f_0$, as continuous increase and step function, respectively.
Further, when the electronic tunnel rates are large compared to $f_0$, near
charge degeneracy points the charge on the quantum dot can adapt (by a fraction
of an elementary charge) to the momentary position within a vibration cycle.
The vibration mode is electrostatically softened \cite{strongcoupling,
nphys-benyamini:151}, cf. Fig.~\ref{fig-gatetrace}(f), proportional to
$\partial \! \left< N \right>\!/\partial\vg$. Thus, resonance frequency minima
indicate the increase of the quantum dot charge $e\!\left\langle N
\right\rangle$ at the charge degeneracy points \cite{strongcoupling,
nphys-benyamini:151}, and $\left\langle N \right\rangle\!(\vg)$ can be
calculated from the frequency evolution $f_0(\vg)$.

\begin{figure}[t]
\includegraphics[width=\columnwidth]{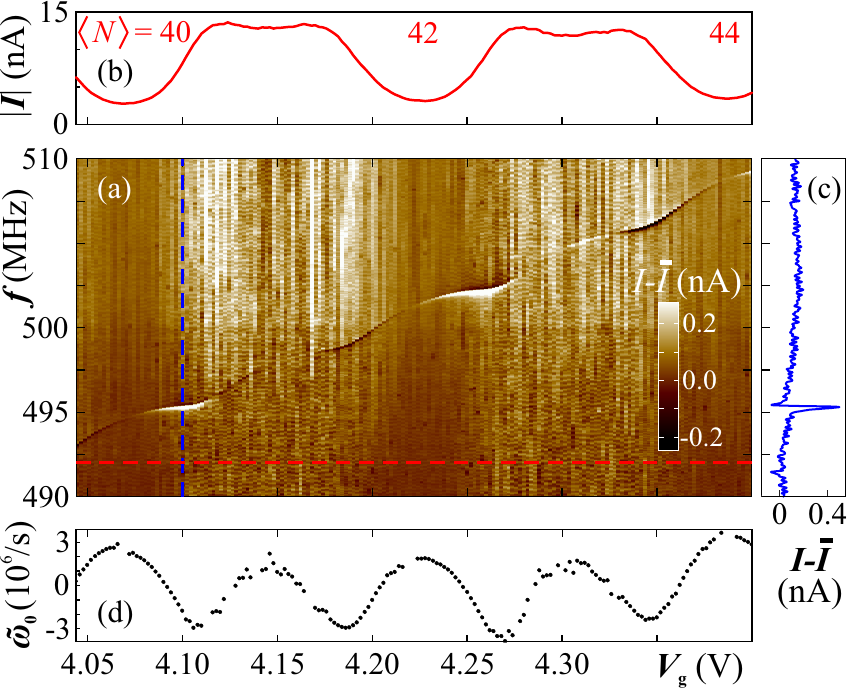}
\caption{\label{fig-resfreq}
(a) Current through the quantum dot, as function of gate voltage \vg\ and rf
driving frequency $f$, with the mean current $\overline{I}(\vg)$ of each
frequency trace subtracted; nominal rf generator power $-25\un{dBm}$, bias
voltage $\vsd =-0.1 \un{mV}$.
(b) $|I(\vg)|$ at off-resonant driving frequency $f=492\un{MHz}$. Kondo enhanced
conductance occurs at odd electron numbers.
(c) Example trace $I(\vg,f) - \overline{I}(\vg)$ from (a) at $\vg=4.1\un{V}$.
The effect of the mechanical resonance on the time-averaged dc current is
clearly visible.
(d) Extracted resonance frequency shift $\tilde{\omega}_{0}(\vg)= 2\pi
f_{0}(\vg)-(a + b \vg)$ with respect to a linear background; see the
Supplement \cite{supplement} for the detailed fit parameters. Device A.
}
\end{figure}
Figure \ref{fig-resfreq}(a) shows a measurement of the vibration-induced signal
in the Kondo regime. For different gate voltages $\vg$ the time-averaged dc
current $I(\vg, f)$ is recorded while sweeping the driving signal frequency
$f$. In Fig.~\ref{fig-resfreq}(a) (and Fig.~\ref{fig-resfreq}(c), which displays
a trace cut from Fig.~\ref{fig-resfreq}(a)), the mean value $\overline{I}(\vg)$
of each frequency sweep has been subtracted for better contrast. The vibration
resonance becomes clearly visible as a diagonal feature. To evaluate its
detailed evolution, we extract $f_0(\vg)$ and plot it in
Fig.~\ref{fig-resfreq}(d) as $\tilde{\omega}_{0}(\vg) = 2\pi f_{0}(\vg)-(a + b
\vg)$, i.e., with a linear background subtracted \footnote{We use the angular
frequency notation to closely follow published literature
\cite{prb-meerwaldt:115454}. See the Supplement \cite{supplement} for the
detailed fit parameters.}. Every single-electron addition into the dot exhibits
a distinct dip. While the off-resonant dc current $I(\vg)$,
Fig.~\ref{fig-resfreq}(b), clearly shows Kondo zero bias conductance anomalies
at odd quantum dot charge \cite{nature-goldhaber:156}, this odd-even electron
number asymmetry is barely visible in the evolution of the resonance frequency
\footnote{As derived in \cite{highq}, the amplitude of the resonance peak
scales with $\text{d}^2 I / \text{d} V_\text{g}^2$, leading to only a weak
resonance signal on the Kondo ridge and a corresponding increased scatter of
the extracted center frequencies there.}.

\begin{figure}[t]
\includegraphics[width=\columnwidth]{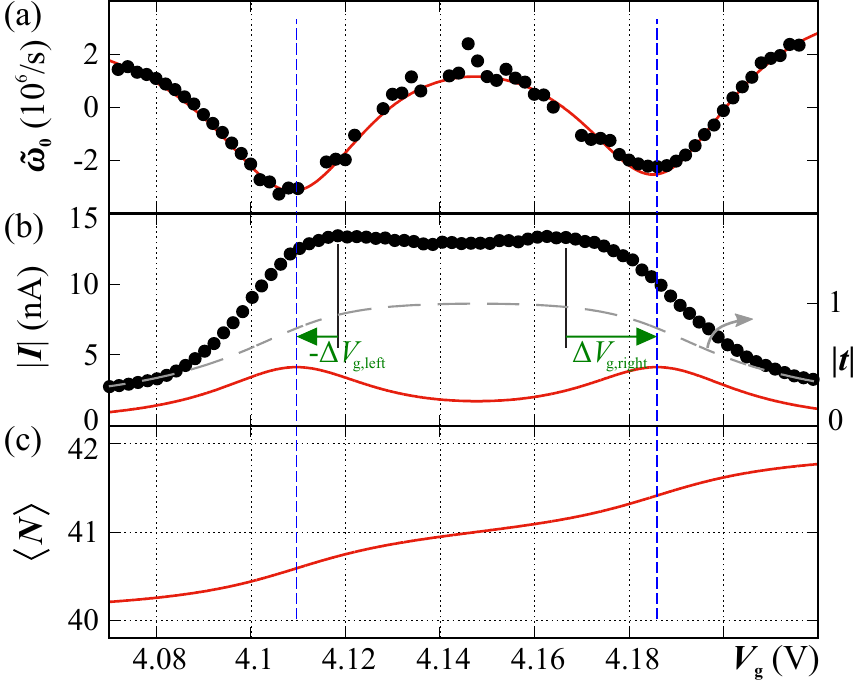}
\caption{\label{fig-chargedetection}
Analysis of the Kondo regime around $\langle N \rangle = 41$;
$\vsd=-0.1\un{mV}$.
(a) Data points: resonance frequency shift $\tilde{\omega}_0(\vg)$, cf.
Fig.~\ref{fig-resfreq}(d). Solid line: curve fit assuming subsequent
occupation of two non-degenerate levels, see the text.
(b) Data points: simultaneously measured off-resonant current $|I|(\vg)$. Solid
red line: sequential tunneling current according to the fit model from (a).
Dashed gray line: $T=0$, $\vsd=0$ Fermi liquid model transmission $\left| t
\right|$ derived from $\langle N \rangle\!(\vg)$ via Friedel's sum rule, see the
text; right axis.
(c) Time-averaged quantum dot occupation $\langle N \rangle\!(\vg)$ derived from
the fit in (a).
}
\end{figure}
{\em Evolution with \vg{}--} 
In Fig.~\ref{fig-chargedetection}(a), we show a detail of the resonance
frequency evolution from Fig.~\ref{fig-resfreq}(d), accompanied by the current
$I(\vg)$ in Fig.~\ref{fig-chargedetection}(b). To model it, we reduce the
quantum dot to two non-degenerate Lorentz-broadened levels, separated by a
capacitive addition energy $U > \Gamma$, without taking any higher order
tunneling effects into account. We only consider the case of linear response,
i.e., $e\vsd \ll \hbar \Gamma$; in addition, for the large transparency of the
contact barriers present at $\nel \approx 41$ and for electron temperatures of
roughly $T\lesssim 50\un{mK}$, we neglect the thermal broadening of the Fermi
distribution in the contacts. Then, $\langle N \rangle\! (\vg)$ is only smeared
out by the lifetime broadening $\Gamma$ of the quantum dot states. The tunnel
barrier transmittances between dot and leads are assumed to be
energy-independent and equal; the tunnel rates $\Gamma^{\pm}_{1/2}$ onto and
off the quantum dot levels are obtained by integrating over the density of
states on the dot.

We use this model to fit the functional dependence of the resonance frequency
to the data in Fig.~\ref{fig-chargedetection}. Following \cite{strongcoupling,
prb-meerwaldt:115454}, the decrease of the resonance frequency at finite single
electron tunneling (cf. Fig.~\ref{fig-gatetrace}(e)) is given by
\begin{equation} 
\Delta \omega_0=\frac{\vg(\vg-V_\text{CNT})}{2m 
\omega_0 \ctot}
\left(\frac{\text{d}\cg}{\text{d}z}\right)^2\left(1
-\frac{e}{\cg}\frac{\partial \left\langle N\right\rangle}{\partial \vg}\right), 
\label{FreqDep}
\end{equation}
with $V_{\text{CNT}}=(\cg \vg - e\left\langle N\right\rangle)/\ctot$ as the
voltage on the CNT, $m$ the nanotube mass, and $\omega_0=2\pi f_0$. The gate
and total capacitances $\cg$ and $\ctot$ are extracted from Coulomb blockade
measurements. Since we do not know the precise position of our CNT, we treat
the capacitive displacement sensitivity $\text{d} \cg/\text{d} z$, where $z$ is
the deflection of the nanotube, as a free parameter. A detailed discussion of
the fit procedure, a table of the device parameters entering the calculation,
and the resulting fit parameters can be found in the Supplement
\cite{supplement}. Note that the relevant gate dependent term in
Eq.~(\ref{FreqDep}) is the quantum capacitance, i.e., the derivative of the
charge occupation, ${\partial\! \left\langle N \right\rangle} / {\partial
\vg}$, also called compressibility in \cite{Desjardins2017}.

Our simplified model reproduces the functional dependence of the resonance
frequency in Fig.~\ref{fig-chargedetection}(a) very well. The result can be
used to derive the expected sequential-tunneling current from our model and the
time-averaged charge evolution $\left< N \right>\! (\vg)$ in the quantum dot,
see the solid lines in Fig.~\ref{fig-chargedetection}. While Kondo processes
absent in our model strongly contribute to electronic transport, they do
not significantly influence the time-averaged occupation of the quantum dot and
thereby the mechanical resonance. This is in excellent agreement with results
by Sprinzak {\it et al.} \cite{prl-sprinzak:176805}, combining a quantum point
contact as charge detector \cite{prl-field:1311, anticrossing} with a
gate-defined quantum dot, as well as recent data analyzing the charge
compressibility of a quantum dot by means of a coupled coplanar waveguide
cavity, see \cite{Desjardins2017}. The suppression of quantum dot charging by
Coulomb blockade is independent of the Kondo enhanced conductance via {\it
virtual} occupation.

{\em Gate potential of current and compressibility maxima--}
In a naive analogy, one would expect that in the Kondo case, as in the case of
strong Coulomb blockade \cite{strongcoupling, prb-meerwaldt:115454,
Desjardins2017}, the increase of the time-averaged charge on the quantum dot
takes place predominantly at the gate voltage of the current maxima. The data
points of Fig.~\ref{fig-chargedetection}(b) show the current $I(\vg)$ at fixed
bias, recorded simultaneously with the mechanical resonance frequency,
Fig.~\ref{fig-chargedetection}(a). Comparing the extrema of the resonance
frequency $\tilde{\omega}_0(\vg)$, Fig.~\ref{fig-chargedetection}(a), and
the current $|I|(\vg)$, Fig.~\ref{fig-chargedetection}(b), distinct shifts
$\Delta V_\text{g,left}$ and $\Delta V_\text{g,right}$ are observed, see the
green arrows.

In experimental literature, a temperature-induced shift of the current maximum
due to Kondo correlations has already been reported in the first publications
\cite{prl-goldhaber:5225}. In the data of Sprinzak {\it et al.},
\cite{prl-sprinzak:176805}, a systematic shift between current and quantum
capacitance extrema similar to our observations is visible (though not
discussed). This confirms that the phenomenon is intrinsic to the Kondo effect
in a quantum dot, independent of the experimental realization. Early
calculations by Wingreen and Meir, \cite{Wingreen1994}, using the noncrossing
approximation in the Anderson model, have already predicted a temperature
dependent shift of the current maximum position (see Fig.~6 and Fig.~7(a) in
\cite{Wingreen1994}).

\begin{figure}[t]
\includegraphics[width=\columnwidth]{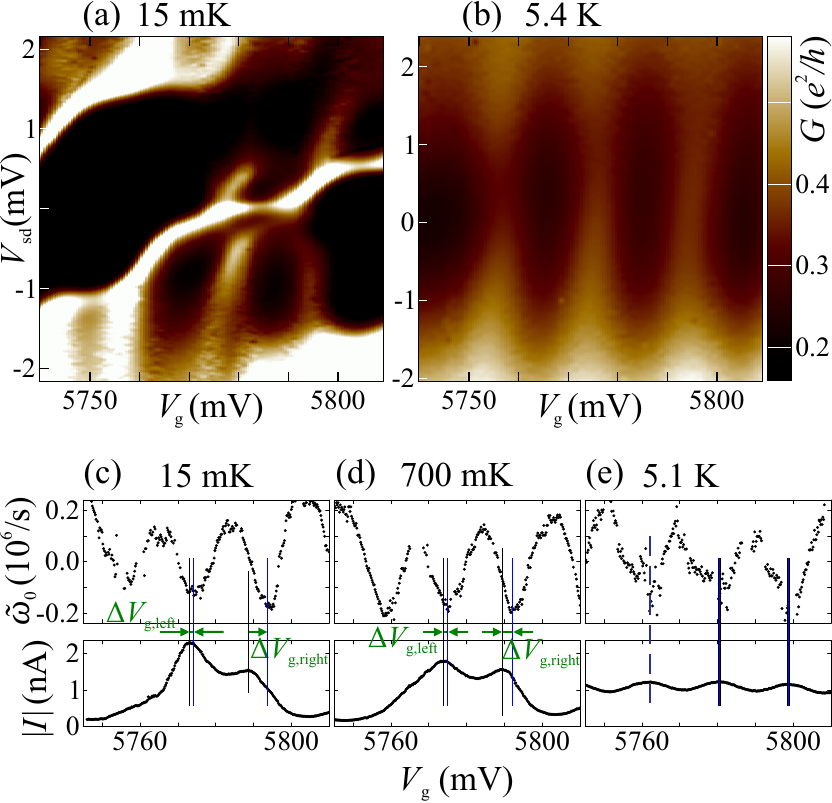}
\caption{\label{fig-temp}
(a), (b) Differential conductance of device B at (a) base temperature $T =
15\un{mK}$ and (b) $T = 5.4\un{K}$. While transport in the millikelvin regime
is dominated by higher order effects, above $T=5\un{K}$ regular, strongly
broadened Coulomb blockade oscillations emerge.
(c--e) Combined plots of mechanical resonance shift
$\tilde{\omega}_{0}(\vg)$ and dc current $|I|(\vg)$, for $\vsd=-0.1\un{mV}$ and
(c) $T=15\un{mK}$, (d) $0.7\un{K}$, (e) $5.1\un{K}$.}
\end{figure}
{\em Temperature dependence--} Figure~\ref{fig-temp} illustrates the
suppression of correlation effects at elevated temperature. In the region of
the figure we obtain Kondo temperatures in the range $1\un{K} \lesssim
T_{\text{K}} \lesssim 5\un{K}$. While the large dot-lead coupling strongly
distorts the stability diagram at base temperature \cite{prl-makarovski:066801}
\footnote{The observed complex transport spectrum of device B in this parameter
range goes beyond the SU(2) Kondo model.}, see Fig.~\ref{fig-temp}(a), at
$T\gtrsim 5\un{K}$ in Fig.~\ref{fig-temp}(b) regular, thermally broadened
Coulomb blockade oscillations reemerge. Figures~\ref{fig-temp}(c-e) display
both extracted mechanical resonance frequency and measured dc current for (c)
$T=15\un{mK}$, (d) $T=0.7\un{K}$, and (e) $T=5\un{K}$. With increasing
temperature the mechanical resonance broadens \cite{highq} and the
determination of the resonance frequency becomes more challenging. At the same
time, the current evolves from a complex, Kondo- and level renormalization
dominated behavior to broadened but regular and in the plotted range
nearly bias-independent Coulomb blockade oscillations.

As expected, at higher temperature charging and current maxima coincide better.
\begin{figure}[t]
\includegraphics[width=\columnwidth]{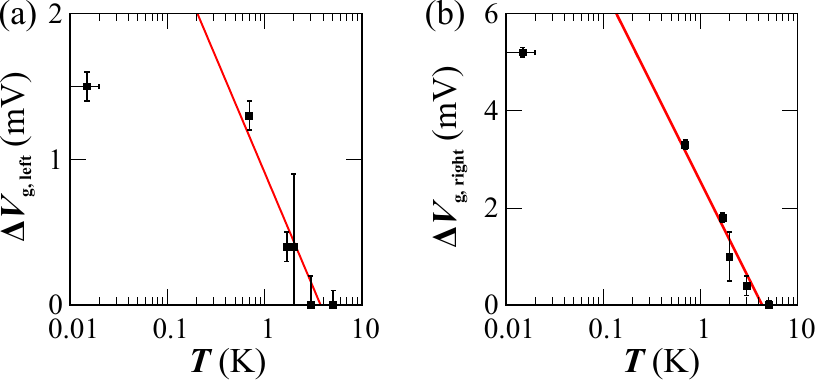}
\caption{\label{fig-tshift}
Temperature dependence of (a) $\Delta V_{\mathrm{g,left}}$ and (b) $\Delta
V_{\mathrm{g,right}}$, the shift between mechanical resonance frequency minimum
and dc current maximum, for the oscillations marked in Fig.~\ref{fig-temp}. The
solid lines correspond to a logarithmic fit, with the $T=15\un{mK}$ point
omitted \cite{Wingreen1994}.
}
\end{figure}
This is quantified in Fig.~\ref{fig-tshift}, where the relative shifts in gate
voltage $\Delta V_{\mathrm{g,left}}$ and $\Delta V_{\mathrm{g,right}}$ between
resonance frequency minimum and current maximum are plotted for two exemplary
Coulomb oscillations marked in Fig.~\ref{fig-temp}. Starting from about $1.5
\un{mV}$, respectively $5.2 \un{mV}$, the peak shifts decrease with increasing
temperature asymptotically towards zero. The straight lines in the figure,
fits excluding the $T=15\un{mK}$ point due to likely saturation there,
correspond to the typical logarithmic scaling present in Kondo phenomena and
predicted for the peak shift \cite{Wingreen1994} and are consistent with the 
data.

{\em Relation to the transmission phase--}
In an early theoretical work on Kondo physics, Gerland {\it et al.}
\cite{prl-gerland:3710} discuss the electronic transmission phase of a Kondo
quantum dot, a topic of intense attention over the previous decades. Friedel's
sum rule \cite{Friedel1956, Langreth1966} intrinsically relates the
transmission phase to the number of electronic states below the Fermi energy
and thereby the time-averaged occupation. This means that we can directly
compare the combined Figs.~3(c) and 3(d) of \cite{prl-gerland:3710}
(transmission magnitude and phase) with our data of
Figs.~\ref{fig-chargedetection}(b) and \ref{fig-chargedetection}(c) here
(current and time-averaged occupation). Indeed, a highly similar functional
dependence is visible; see the Supplement \cite{supplement} for a detailed comparison. With this
background and based on Fermi-liquid theory of the SU(2) Kondo effect, the
dashed gray line in Fig.~\ref{fig-chargedetection}(b) plots the transmission
amplitude evolution $\left| t(\vg) \right|=\sin (\pi \left< N
\right>\!(\vg)/2)$ of the quantum dot expected for $\vsd=T=0$. This clearly
demonstrates the Kondo ridge as well as the distinct shift between large
transmission magnitude and maximum slope of the transmission phase. The
deviations in current behaviour $I(\vg)$ may be due to the finite temperature
and bias, and/or indicate an experimental situation more complex than the
SU(2) Kondo effect.

{\em Conclusion--} We use the mechanical resonance frequency of a suspended
carbon nanotube to trace the average electronic occupation of a strongly
Kondo-correlated quantum dot embedded in the nanotube. We show that sequential
tunneling alone already provides a good model for the average charge $\langle N
\rangle(\vg)$ and the mechanical resonance frequency $\omega_0(\vg)$. While
dominant for electronic transport (conductance), the influence of Kondo
correlations on the time averaged charge and thereby the mechanical system is
small in the chosen parameter regime. We observe a distinct shift in gate
voltage of the current maxima, relative to the maxima of the charge
compressibility $\partial \langle N \rangle/\partial \vg$, effectively
distorting the Coulomb blockade regions. This shift decays with increasing
temperature, a clear signature that it is caused by the Kondo correlations. Our
results are in excellent agreement with theoretical modelling
\cite{prl-gerland:3710, Wingreen1994}.

Future work, applying our highly versatile sensing method to higher harmonic
modes of the vibration, may address the parameter region $f_\text{mech} > k_B
T_K$ \cite{nl-laird:193, prb-delagrange-2018}, or even the charge distribution
along the carbon nanotube axis via a spatially modulated electron-vibration
coupling \cite{nphys-benyamini:151}. Kondo phenomena in carbon nanotubes beyond
the SU(2) spin Kondo effect provide further systems of obvious experimental and
theoretical interest.

\begin{acknowledgments}
The authors acknowledge financial support by the Deutsche 
Forschungsgemeinschaft (Emmy Noether grant Hu 1808/1, GRK 1570, SFB 689) and by 
the Studienstiftung des deutschen Volkes. We thank J. von Delft, J. Kern, A.
Donarini, M. Marga\'nska, and M. Grifoni for insightful discussions.
\end{acknowledgments}

\end{document}